\documentclass[superscriptaddress,preprintnumbers,amssymb,twocolumn]{revtex4}
\usepackage{graphicx}
\usepackage[latin1]{inputenc}
\usepackage{epsfig}
\usepackage{amsmath}
\usepackage{amsfonts}
\usepackage{amssymb}

\def\lsim{\:\raisebox{-0.5ex}{$\stackrel{\textstyle<}{\sim}$}\:}
\def\gsim{\:\raisebox{-0.5ex}{$\stackrel{\textstyle>}{\sim}$}\:}

\begin{document}

\title {Hybrid dynamical electroweak symmetry breaking with heavy quarks and the 125 GeV Higgs}
\author{Michael Geller}
\email{mic.geller@gmail.com}
\affiliation{Physics Department, Technion-Institute of Technology, Haifa 32000, Israel}
\author{Shaouly Bar-Shalom}
\email{shaouly@physics.technion.ac.il}
\affiliation{Physics Department, Technion-Institute of Technology, Haifa 32000, Israel}
\author{Amarjit Soni}
\email{adlersoni@gmail.com}
\affiliation{Theory Group, Brookhaven National Laboratory, Upton, NY 11973, USA}

\date{\today}

\begin{abstract}
Existing  models of dynamical electroweak symmetry breaking (EWSB)
find it very difficult to get a Higgs of
mass lighter than $m_t$. Consequently, in light of the LHC discovery of
the $\sim 125$ GeV Higgs, such models face a significant
obstacle.  Moreover, with three generations those models
have a superheavy cut-off around $10^{17}$ GeV, requiring a
significant fine-tuning.
To overcome these twin difficulties, we
propose a hybrid framework for EWSB,
in which the Higgs mechanism is combined with a
Nambu-Jona-Lasinio mechanism.
The model introduces a strongly coupled doublet of heavy
quarks with a mass around 500 GeV, which forms a condensate
at a compositeness scale
$\Lambda$ about a few TeV, and an additional
unconstrained scalar doublet which behaves as a "fundamental"
doublet at $\Lambda$. This "fundamental"-like doublet
has a vanishing quartic term at $\Lambda$ and is, therefore, not
the SM doublet, but should rather be viewed as a pseudo-Goldstone boson
of the underlying strong dynamics.
This setup is matched at the compositeness scale $\Lambda$
to a tightly constrained hybrid two Higgs doublet model, where
both the composite and unconstrained scalars
participate in EWSB.
This allows us
to get a good candidate for the recently observed 125 GeV scalar
which has properties very similar to the Standard Model Higgs.
The heavier (mostly composite) CP-even scalar has a mass
around 500 GeV, while the pseudoscalar and the charged Higgs particles
have masses in the range $200 -300$ GeV.
\end{abstract}


\maketitle

\section{Introduction}

With the recent LHC discovery of the 125 GeV scalar particle
\cite{LHCHiggs},
we are one step closer to understanding the mechanism of EWSB.
The Standard Model (SM) Higgs mechanism lacks a fundamental
explanation and has, therefore, been long questioned.
Indeed, inspired from our experience with QCD, it has been speculated
that the scalar responsible for EWSB may not
be fundamental but, instead, some form of a fermion-antifermion bound state, which is
generated dynamically by some new strong interaction at a higher scale \cite{EWSB-papers}.

The ``conventional" models for this scenario
are based on the
Nambu-Jona-Lassinio (NJL) mechanism \cite{Nambu}, in which Dynamical EWSB (DEWSB) is triggered by the condensation of heavy fermions.
However, such models have one major caveat:
the typical mass of the
heavy fermionic condensate, $H \sim < \bar\Psi \Psi >$,
tends to lie in the range
$m_\Psi < m_H < 2m_\Psi$ (see e.g., \cite{Bardeen}).
Thus,
such a composite tends to be too heavy to account for the recently
discovered 125 GeV Higgs-like particle.$^{[1]}$\footnotetext[1]{One
possible way out, which we will not consider here,
is that the lightest scalar state is the pseudoscalar associated
with DEWSB (see e.g., \cite{DEWSB-Pseudo}),
since its mass does not receive large corrections from loops of
the heavy fermions and, thus, can in principle be held small without
fine-tuning.} Because of this difficulty, in this paper we propose an alternative solution
for the TeV-scale DEWSB scenario,
in which a light SM-like Higgs with a mass of ${\cal O}(m_W)$ emerges.
We basically construct, as we will elaborate later,
a hybrid DEWSB setup by adding an unconstrained scalar field at the
compositeness scale, which behaves essentially like a ``fundamental" field.

Indeed, one of the early attempts in this direction investigated
the possibility of using the top-quark
as the agent of DEWSB via top-condensation
\cite{Bardeen}, in a generalization
of the NJL model.
However, the resulting dynamical top mass turns out to be
appreciably heavier than $m_t$,
thus making it difficult for top condensation to provide a viable picture.
Moreover, top-condensate
models [or NJL models where the condensing fermions have masses
of ${\cal O}(m_t)$], require the cutoff
for the new strong interactions to be many orders of
magnitudes larger than $m_t$, i.e., of ${\cal O}(10^{17})$ GeV,
resulting in a severely fine-tuned picture of DEWSB.

In passing,  we should mention that
several interesting generalizations of the top-condensate
model of \cite{Bardeen},
which potentially avoid these obstacles, have been suggested. For example,
one can relax the requirement that only the top-condensate is responsible
for the full EWSB \cite{Hill1994,newpaper}, or assume
that condensations of new heavier quarks
and/or leptons drive EWSB \cite{DEWSB-heavyF,luty,DEWSB-HLP,burdman1,hunghybrid}.
In such scenarios
the resulting low-energy (i.e., EW-scale) effective
theory may contain more than a single composite Higgs doublet
\cite{newpaper,luty,DEWSB-HLP,burdman1,hunghybrid,DEWSB-multiH,DEWSB-Pseudo}.
Indeed, low-energy multi-Higgs models,
with new heavy fermions with masses of ${\cal O}(500~{\rm GeV})$,
are natural outcomes of a TeV-scale DEWSB scenario, since
the heavy fermions are expected to be
strongly coupled at the near by TeV-scale and to lead to
the formation of several condensates - possibly with sub-TeV masses.

In this paper, as alluded to above, we take another direction constructing
a hybrid DEWSB setup by adding an unconstrained
scalar field
at the compositeness scale which behaves as a ``fundamental" field
where additional
super-critical attractive 4-Fermi operators form
a composite scalar sector. The ``fundamental" scalar is unconstrained
at the compositeness scale and may
result from the underlying strong dynamics \cite{Hill1994},
e.g., it can be the pseudo-Goldstone boson of a global symmetry
breaking at the strong interaction scale.
This strongly coupled composite-plus-fundamental sectors
are matched at the compositeness scale
to a hybrid ``4th generation" 2HDM (h4G2HDM),
with one fundamental-like field ($\Phi_\ell$) which couples
to the SM's light fermions and
one auxiliary (composite) field ($\Phi_h$)
which couples to the heavy
quarks.$^{[2]}$\footnotetext[2]{Other interesting hybrid multi-Higgs
model were suggested in \cite{hunghybrid,hybrid-top}. In particular, in
\cite{hunghybrid} the condensates which drive DEWSB
are formed by exchanges of the
fundamental Higgs, while in \cite{hybrid-top} a hybrid setup similar to ours
was constructed in the top-condensation scenario.}
The fundamental or unconstrained Higgs field
is thus responsible for the mass generation of the lighter SM fermions
and for the observed CKM flavor pattern.

We stress that the name ``4th generation" is
used here for convenience only and should not be confused with the minimal SM4
framework, as we explicitly involve a 2HDM.
Furthermore, the DEWSB mechanism proposed here can be
generalized to the case of
non-sequential heavy quarks, e.g.,
new heavy vector-like quarks \cite{vectorquarks}
(see further discussion in the summary).
Later in this paper, we will briefly explore the phenomenology and study
the confrontation of this model with the latest LHC data as well as
flavor and EW precision constraints.

Our h4G2HDM setup is motivated by the concept that
new heavy fermions
are expected to have purely dynamical masses,
while a different mechanism is expected to underly
the generation of mass for the lighter SM fermions.
It further
allows us to get a light 125 GeV SM-like Higgs,
since its mass is mostly proportional
to the quartic coupling of the fundamental field,
which is unconstrained at the compositeness scale.
That is, with this assignment, in our h4G2HDM
the mass of the lightest CP-even Higgs state
does not receive the usual large quantum
corrections from loops of the heavy
dynamical fermions, which instead feed
into the quartic coupling of the composite scalar.

It should be noted that the simplest
low-energy effective setup which may result
from a TeV-scale DEWSB scenario
is the so called SM4, i.e.,
the SM with a 4th sequential generation of
heavy fermions and one Higgs doublet.
This minimal framework has, however,
several drawbacks \cite{4G2HDM1} and, more importantly,
it fails to account for
the existence of the recently discovered
125 GeV Higgs-like particle \cite{SM4-excluded}.
On the other hand, a 4th generation
framework with two (or more) Higgs doublets,
such as the h4G2HDM discussed here,
where the new heavy quarks have masses of ${\cal O} \left(500~ {\rm GeV} \right)$,
is not only consistent with the current Higgs data$^{[3]}$\footnotetext[3]{For
discussions on the phenomenology of multi-Higgs 4th
generation models, see e.g., \cite{DEWSB-multiH,4G2HDM1,4G2HDM2,4G2HDM3,4G2HDM4,4G2HDMpheno}.}, but
it can also
have specific flavor structures, which
may give rise to new signatures of 4th generation quarks
at the LHC \cite{4G2HDM1,4G2HDM2}, and
substantially relax the current bounds on their masses  \cite{4G2HDM3}.
Moreover, the lightest Higgs state in these class of models may be a good candidate for the
recently discovered 125 GeV Higgs-like
particle \cite{4G2HDM4}.

\section{Dynamical EW symmetry breaking with heavy fermions: a Hybrid setup}

We assume that the 4th generation quarks are charged under
some new strong interaction that dynamically breaks
EW symmetry (see e.g., the ``top-color" models of \cite{Hill1994,Hill1991}).
The theory at the compositeness scale,
$\Lambda$, can then be parameterized
by adding to the light SM degrees of freedom the following
set of strongly coupled 4-Fermi terms:
\begin{eqnarray}
{\cal L}={\cal L}_{SM}(\Lambda) &+& G_{t^\prime} \bar Q_L^\prime t^\prime_R {\bar t}^\prime_R Q_L^\prime +
G_{b^\prime} \bar Q_L^\prime b^\prime_R {\bar b}^\prime_R Q_L^\prime \nonumber \\
&+& G_{t^\prime b^\prime} \left( \bar Q_L^\prime b^\prime_R {\bar t}^{\prime c}_R i \tau_2 Q_L^{\prime c} + h.c. \right) ~,
\label{4fermi}
\end{eqnarray}
where $Q_L^\prime = (t^\prime_L ~ b^\prime_L)^T$ and ${\cal L}_{SM}(\Lambda)$
stands for the bare SM
Lagrangian with a single
fundamental Higgs field, $\Phi_\ell$,
which essentially parameterizes our ignorance as for
the origin of mass for the lighter fermions. It is given by:
\begin{eqnarray}
{\cal L}_{SM}(\Lambda) &=& \left| D_\mu \Phi_\ell \right|^2 + {\cal L}_{SM}^Y(\Lambda)
- V_{SM}(\Lambda) ~, \nonumber \\
{\cal L}_{SM}^Y(\Lambda)  &=& g_{u}^{0,ij} \bar Q_L^i \tilde\Phi_\ell u^j_R
+ g_{d}^{0,ij} \bar Q_L^i \Phi_\ell d^j_R + h.c. ~, \nonumber \\
V_{SM}(\Lambda) &=&
(\mu_\ell^0)^2 \Phi_\ell^\dagger \Phi_\ell +
\frac{1}{2}  \lambda_\ell^0 \left( \Phi_\ell^\dagger \Phi_\ell \right)^2 ~,
\label{SMphi}
\end{eqnarray}
where $\tilde\Phi \equiv i \tau_2 \Phi^\star$, $i,j=1-3$ and
we use the superscript 0 to denote bare couplings at the scale $\Lambda$ .

We can reproduce the theory defined by Eq.~\ref{4fermi} by introducing at $\Lambda$
an auxiliary Higgs doublet, $\Phi_h$, which couples only to the 4th generation quarks as
follows:
\begin{eqnarray}
{\cal L}_{q^\prime}(\Lambda) &=&
g_{b^\prime}^0 \left( \bar Q_L^\prime \Phi_h b^\prime_R + h.c. \right) +
g_{t^\prime}^0 \left( \bar Q_L^\prime \tilde\Phi_h t^\prime_R + h.c. \right) \nonumber \\
&-& (\mu_h^0)^2 \Phi_h^\dagger \Phi_h ~.
\label{aux1}
\end{eqnarray}

For simplicity we did not write above the Yukawa
terms for the light and for the 4th generation leptons. In particular,
our results are not sensitive to the choice by which the 4th generation
leptons couple to the Higgs sector; they can either couple
to the fundamental Higgs or to the auxiliary field.
In either case, we assume that
their couplings are sub-critical and, therefore,
do not play any role in DEWSB (see also discussion below).

The scalar sector of the full theory at the compositeness scale is, therefore, described
by:
\begin{eqnarray}
{\cal L}(\Lambda) = {\cal L}_{SM}(\Lambda) +
{\cal L}_{q^\prime}(\Lambda)
+ (\mu_{h \ell}^0)^2 \left( \Phi_h^\dagger \Phi_\ell + h.c. \right) ~,
\label{aux2}
\end{eqnarray}
where we have added a $\Phi_h - \Phi_\ell$ mixing term
$ (\propto \mu_{h \ell}^0)^2$, which may arise e.g., from QCD-like instanton effects
associated with the underlying strong dynamics (see e.g., \cite{Hill1994,PQ1})
or from sub-critical couplings of the
fundamental Higgs to the 4th generation quarks (see below). This term
explicitly breaks the $U(1)$ Peccei-Quinn (PQ) symmetry \cite{PQ},
which is otherwise possessed by the model, thus
avoiding the presence of a massless pseudoscalar in the spectrum.
Note that, in any realistic scenario we expect
$\mu_{h \ell}(\mu \sim m_W) \sim {\cal O}(m_W)$ and,
since this is the only term which breaks the
PQ symmetry, it evolves only logarithmically under the RGE
so that, at the compositeness scale, we also have
$\mu_{h\ell}^0 \equiv \mu_{h \ell}(\mu \sim \Lambda) \sim {\cal O}(m_W)$.
Therefore, since
$\mu_{h/\ell}^0 \equiv \mu_{h/\ell}(\mu \sim \Lambda) \sim {\cal O}(\Lambda)$,
we expect $(\mu_{h \ell}^0)^2/(\mu_{h}^0)^2 \sim {\cal O}(m_W^2/\Lambda^2) \ll 1$.

When the auxiliary (composite) field is integrated out at
$\Lambda$, we recover the Lagrangian defined by Eq.~\ref{4fermi}, with
\begin{eqnarray}
G_{t^\prime} = \frac{(g_{t^\prime}^0)^2}{(\mu_h^0)^2}  ~,~
G_{b^\prime} = \frac{(g_{b^\prime}^0)^2}{(\mu_h^0)^2}  ~,~
G_{t^\prime b^\prime} = -\frac{g_{t^\prime}^0 g_{b^\prime}^0}{(\mu_h^0)^2} ~,
\label{Gterms}
\end{eqnarray}
plus additional interaction terms between the light Higgs and the new heavy quarks
with Yukawa couplings of
${\cal O} \left( \frac{(\mu_{h \ell}^0)^2}{(\mu_{h}^0)^2} \cdot g_{t^\prime/b^\prime}^0 \right)$;
since at $\Lambda$ we have $(\mu_{h \ell}^0)^2/(\mu_{h}^0)^2 \sim {\cal O}(m_W^2/\Lambda^2) \ll 1$,
such residual $\Phi_\ell \bar q^\prime q^\prime$ terms are expected to be
small and will, therefore, not participate in EWSB.

Thus, from the point of view of DEWSB,
the hybrid 2HDM defined at $\Lambda$, i.e.,
with one fundamental and one auxiliary/composite scalar fields, is exactly equivalent
to the theory defined in Eq.~\ref{4fermi} with the strong 4-Fermi interactions
of the 4th generation quarks.
The auxiliary field, $\Phi_h$, is, therefore, viewed as a composite
of the form
$\Phi_h \sim g_{t^\prime}^\star < \bar Q_L^{\prime c} (i \tau_2) t^{\prime c}_R > + g_{b^\prime} < \bar Q_L^\prime b^\prime_R >$,
which is responsible
for EWSB and for the dynamical mass generation of the heavy quarks.
At low energies the field $\Phi_h$ acquires a kinetic term as well
as self interactions
and the theory behaves as a 2HDM with a structure similar to the
4G2HDM proposed in \cite{4G2HDM1}, i.e., one Higgs field ($\Phi_h$) couples
only to the heavy 4th generation quarks and the
2nd Higgs field ($\Phi_\ell$)
couples to the SM quarks of the 1st-3rd generations.
The mass terms $\mu_{h},\mu_{\ell}$ receive quantum corrections,
resulting in EW-scale VEV's for $\Phi_h$ and for
$\Phi_\ell$ which break the EW symmetry.

The resulting low-energy h4G2HDM scalar potential can be written as:
\begin{widetext}
\begin{eqnarray}
V_{h4G2HDM}(\Phi_h,\Phi_\ell)  &=& \mu_\ell^2 \Phi_\ell^\dagger \Phi_\ell + \mu_h^2 \Phi_h^\dagger \Phi_h
-  \mu_{h \ell}^2 \left( \Phi_h^\dagger \Phi_\ell + h.c. \right) \nonumber \\
&+& \frac{1}{2} \lambda_\ell \left( \Phi_\ell^\dagger \Phi_\ell \right)^2 +
\frac{1}{2} \lambda_h \left( \Phi_h^\dagger \Phi_h \right)^2 +
\lambda_3 \left( \Phi_h^\dagger \Phi_h \right) \left( \Phi_\ell^\dagger \Phi_\ell \right)
+ \lambda_4 \left( \Phi_h^\dagger \Phi_\ell \right) \left( \Phi_\ell^\dagger \Phi_h \right)~.
\label{V}
\end{eqnarray}
\end{widetext}
where all the above mass terms and quartic couplings run as a function
of the energy scale $\mu$, as dictated
by the RGE for this model.$^{[4]}$\footnotetext[4]{The most
general 2HDM potential also includes the
quartic couplings $\lambda_{5,6,7}$ \cite{branco}, which, in our
h4G2HDM, are absent at any scale.}
The stability condition for the above potential
reads $\lambda_\ell,\lambda_h > 0$ and
$\sqrt{\lambda_\ell \lambda_h} > -\lambda_3 - \lambda_4$.
Also, we apply the compositeness boundary conditions
at the scale $\Lambda$ to
the strong 4th generation Yukawa couplings
that generate the 4-Fermi terms
and to the quartic couplings involving the auxiliary field $\Phi_h$:
\begin{eqnarray}
&& g_{q^\prime}(\Lambda) \to \infty ~~,~~ \lambda_{h,3,4}(\Lambda) \to \infty ~, \nonumber \\
&& \lambda_h(\Lambda)/g_{q^\prime}^4(\Lambda) \to 0 ~~,~~
\lambda_{3,4}(\Lambda)/g_{q^\prime}^2(\Lambda) \to 0 ~, \label{cc}
\end{eqnarray}
where $q^\prime = t^\prime,b^\prime$. Eq.~\ref{cc} reflects
the fact that the h4G2HDM is matched at $\Lambda$ to the
theory defined by Eq.~\ref{4fermi},
with the strongly coupled 4-Fermi factors
derived in Eq.~\ref{Gterms}.
On the other hand, the quartic coupling of the
fundamental Higgs is unconstrained at $\Lambda$, so that we
have $\lambda_\ell (\mu \to \Lambda) \to \lambda_\ell^{(0)}$,
where $\lambda_\ell^{(0)}$ is a free parameter of the model.

The boundary conditions for $\mu_{h},\mu_\ell$ and $\mu_{h\ell}$
are not required for our analysis, since we are only interested
in their values at the EW-scale, which
are fixed by the minimization conditions of the scalar-potential.
In particular, at the minimum of the potential (i.e., at the EW-scale)
we can express $\mu_{h}$ and $\mu_\ell$ in terms of $\mu_{h \ell}$, $\tan\beta \equiv t_\beta = v_h/v_\ell$,
$v = \sqrt{v_h^2 + v_\ell^2}$ and the quartic couplings $\lambda_{h},\lambda_\ell$
(neglecting $\lambda_3$ and $\lambda_4$, see below):
\begin{eqnarray}
\mu_{\ell}^2 \simeq \mu_{h \ell}^2 / t_\beta - v^2 c_\beta^2 \lambda_\ell / 2 ~~,~~
\mu_{h}^2 \simeq t_\beta \mu_{h \ell}^2 - v^2 s_\beta^2 \lambda_h / 2  ~,
\label{muterms}
\end{eqnarray}
where $s_\beta,c_\beta=\sin\beta,\cos\beta$ and it is understood
that the quartic couplings are evaluated at $\mu \sim v$,
i.e., $\lambda_{h} = \lambda_{h}(\mu \sim v)$ and
$\lambda_\ell=\lambda_{\ell}(\mu \sim v)$.

\section{Results and Discussion}

Solving the RGE of the h4G2HDM with the compositeness boundary conditions
in Eq.~\ref{cc},
we find that $d \lambda_{3,4}/d\mu$ is proportional
only to the (small) gauge couplings, so that
$\lambda_3(v),\lambda_4(v) << \lambda_\ell(v),\lambda_h(v)$ and
can therefore be neglected.
Also, in our h4G2HDM, the RGE for $\lambda_\ell$ and for
the top-quark Yukawa coupling $g_t$ are similar
to the SM RGE for
these couplings.
As for the RGE for $\lambda_h$ and for the Yukawa
couplings of the 4th generation quarks $g_{q^\prime}(\mu)$, we
can obtain a viable approximate analytic solution
by
neglecting the contributions from the running of
the gauge couplings and
the Yukawa couplings of all light fermions, as well as
the Yukawa couplings of the 4th generation leptons.
In particular, taking for simplicity
$g_{t^\prime}=g_{b^\prime} \equiv g_{q^\prime}$, the
dominant (approximate) RGE in our h4G2HDM are given by:
\begin{eqnarray}
{\cal D}g_{q^\prime} & \approx & 6 g_{q^\prime}^3  ~, \label{RGE1}\\
{\cal D}\lambda_{h} & \approx & 4 \lambda_{h} \left( 3 \lambda_{h}+ 6 g_{q^\prime}^2 \right) - 24 g_{q^\prime}^4 ~, \label{RGE3}
\end{eqnarray}
where ${\cal D} \equiv 16 \pi^2 \mu \frac{d}{d\mu}$.
With the compositeness boundary conditions of Eq.~\ref{cc}, the above RGE's have a simple analytic solution:
\begin{eqnarray}
g_{q^\prime} \left(\mu\right) = \sqrt{\frac{4\pi^2}{3ln \frac{\Lambda}{\mu}}} ~~~,~~~
\lambda_{h} \left(\mu\right) = \frac{4\pi^2}{3ln \frac{\Lambda}{\mu}} \label{solutions} ~.
\end{eqnarray}

Thus, using $m_{q^\prime} = v_h g_{q^\prime}(\mu =m_{q^\prime})/\sqrt{2}$, we
can obtain
the cutoff $\Lambda$ as a function of $m_{q^\prime}$ and $t_\beta$:
\begin{eqnarray}
\Lambda \approx m_{q^\prime} \cdot
exp \left( \frac{2\pi^2\left( s_\beta v \right)^2}{3 m_{q^\prime}^2} \right)
\label{lambda} ~.
\end{eqnarray}

In particular, for $m_{q^\prime} = 400 - 600$ GeV
and $t_\beta \sim {\cal O}(1)$, for which our low-energy h4G2HDM
successfully accounts for the recently discovered 125 GeV Higgs-like particle (see below),
we find that $\Lambda \sim 1 -1.5$ TeV.
Remarkably this is some fourteen orders of magnitudes smaller than
the cutoff which emerges from a top-condensate scenario:
$\Lambda \sim m_t \cdot exp\left(\frac{16\pi^2v^2}{9m_{t}^2} \right) \sim
10^{17}$ GeV, i.e., obtained by solving the SM-like RGE
for $g_t$: ${\cal D}g_{t} \approx \frac{9}{2} g_{t}^3$.
Thus, introduction of a heavy quark doublet
significantly alleviates the inherent
fine-tuning (i.e., hierarchy) problem that afflicts
the DEWSB models where the condensing fermions have
masses $\lsim 200$ GeV.$^{[5]}$\footnotetext[5]{For an interesting recent
DEWSB model with three generations and
$\Lambda \sim 10^{17}-10^{18}$ GeV, see \cite{newpaper}.}

The physical scalar masses are
$m_A^2 = m_{H^+}^2 = \mu_{h \ell}^2/s_\beta/c_\beta$ and
$m_{h,H}^2 = \left( m_1^2 + m_2^2 \mp \sqrt{(m_1^2 - m_2^2)^2 + 4 \mu_{h \ell}^4} \right)/2 $,
where (see also Eq.~\ref{muterms}):
\begin{eqnarray}
m_1^2 &\simeq&  \mu_{h}^2 + 3 s_\beta^2 v^2 \lambda_h /2 \simeq t_\beta \mu_{h \ell}^2 + s_\beta^2 v^2 \lambda_h ~, \\
m_2^2 &\simeq&  \mu_{\ell}^2 + 3 c_\beta^2 v^2 /2 \lambda_\ell \simeq \mu_{h \ell}^2 / t_\beta +
c_\beta^2 v^2 \lambda_\ell ~, \label{m1m2}
\end{eqnarray}
and a Higgs mixing angle:
\begin{eqnarray}
\tan 2\alpha \simeq \left( \cot 2 \beta - v^2
\left( s_\beta^2 \lambda_h - c_\beta^2 \lambda_\ell \right)/2/\mu_{h \ell}^2 \right)^{-1} \label{mixing}~,
\end{eqnarray}
defined by $h = \cos\alpha \cdot {\rm Re}(\Phi_\ell^0) -
\sin\alpha \cdot {\rm Re}(\Phi_h^0)$ and
$H = \cos\alpha \cdot {\rm Re}(\Phi_h^0) +
\sin\alpha \cdot {\rm Re}(\Phi_\ell^0)$.

Now, since the physical Higgs masses are sensitive to the energy scale
at which the quartic couplings $\lambda_h$ and $\lambda_\ell$ are
evaluated and since we do not apply the Higgs threshold corrections
when solving the RGE,
we need to estimate the errors on $m_h(\mu \sim m_h)$
and on $m_H(\mu \sim m_H)$.
We do so by computing the difference between the masses
obtained with $\lambda_{h/l}(\mu = 100~{\rm GeV})$ and with
$\lambda_{h/l}(\mu= 400~{\rm GeV})$.
We then find that the typical error is of ${\cal O}(\pm 10\%)$  for
$m_h$ and of ${\cal O}(\pm 20\%)$ for $m_H$.
\begin{figure}[htb]
\begin{center}
\includegraphics[scale=0.6]{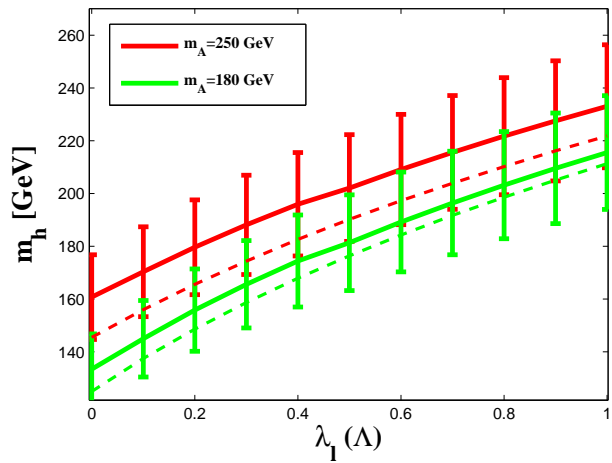}
\end{center}
\caption{\emph{$m_h$ as a function of $\lambda_\ell(\Lambda)$, for
$\tan\beta=0.7$, $m_{q^\prime} = 400$ GeV and two representative values of $m_A$.
The approximate analytic solutions are shown by solid lines and
exact results (obtained from a full RGE analysis, see text)
without errors by the dashed lines.}
\label{fig2}}
\end{figure}

As inputs we use $m_A$ (recall that $\mu_{h \ell} = \sqrt{s_\beta c_\beta} m_A$),
$\tan\beta$, $v=246$ GeV, $m_{q^\prime}$ (which sets the value of $\Lambda$,
see Eq.~\ref{lambda})
and
$\lambda_\ell(\mu \sim \Lambda)$,
while
$\lambda_h(\mu)$ and $g_{q^\prime}(\mu)$ are calculated from
Eq.~\ref{solutions}.
In Fig.~\ref{fig2} we show the typical dependence of $m_h$ on
$\lambda_\ell(\Lambda)$, for $m_{q^\prime} = 400$ GeV and for
some representative values of $\tan\beta$ and $m_A$.
We see that $m_h$ decreases with $\lambda_\ell(\Lambda)$,
so that the minimal $m_h$ is obtained for
$\lambda_\ell(\Lambda)=0$, i.e., at the boundary below which the
vacuum becomes unstable.
Note that, for $\lambda_\ell(\Lambda) \to 0$,
we find $\lambda_\ell(m_W) \sim {\cal O}(0.1)$
and, in particular, $\lambda_h(m_W) \gg \lambda_\ell(m_W)$.
Thus, with $\lambda_\ell(\Lambda) \to 0$
and $\tan\beta \sim {\cal O}(1)$, we obtain
$m_h \sim m_A/\sqrt{2}$ and $m_H \sim v \sqrt{\lambda_h/2}$.
It is also interesting to note that $\lambda_\ell(\Lambda) \to 0$
at the compositeness scale, raises the possibility that the "fundamental"
scalar is a pseudo-Goldstone of some global symmetry breaking in the
underlying strongly interacting theory, as in this case one
expects a zero potential for $\Phi_\ell$ at tree-level
with small loop corrections at energies below $\Lambda$.

\begin{figure}[htb]
\begin{center}
\includegraphics[scale=0.6]{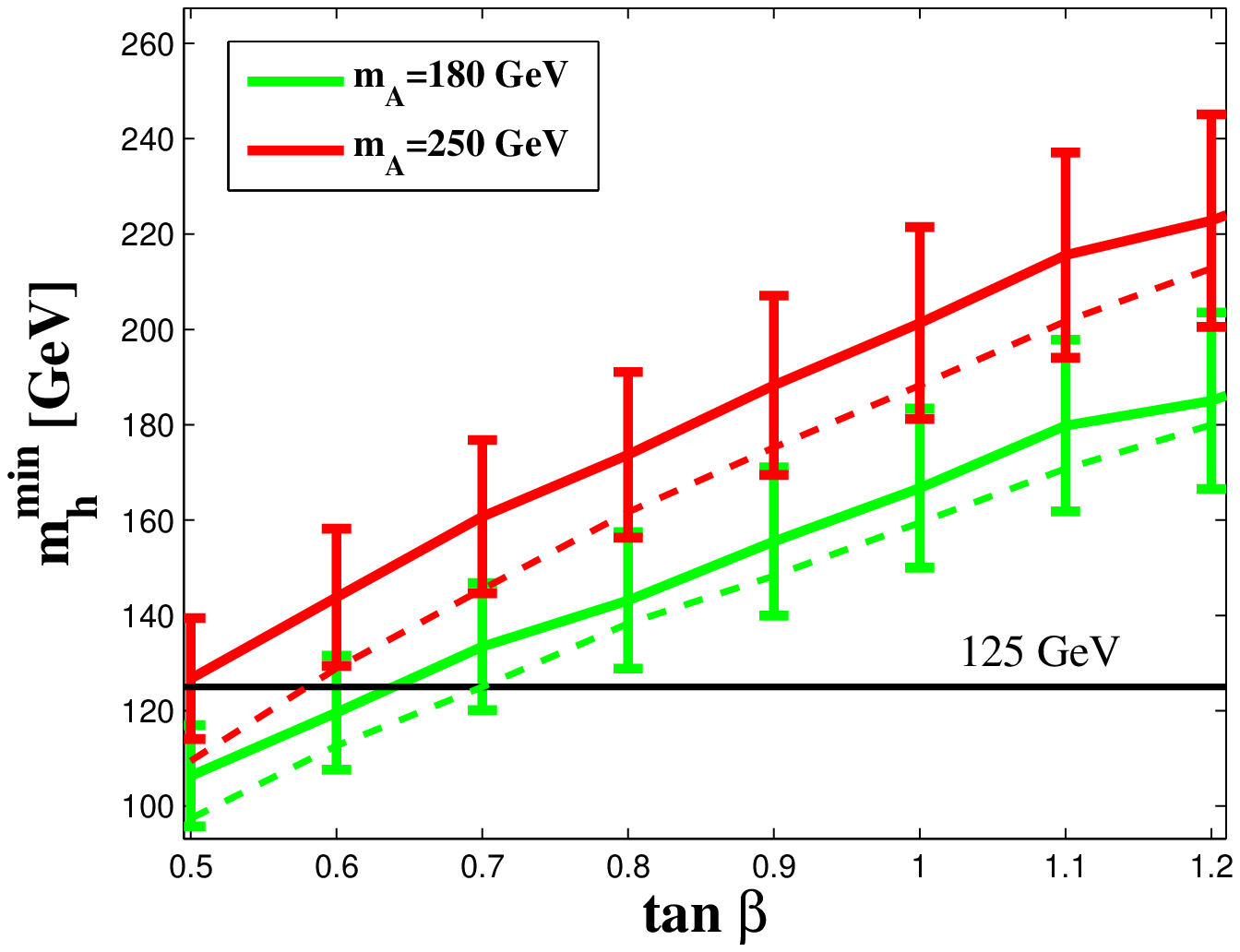}
\includegraphics[scale=0.6]{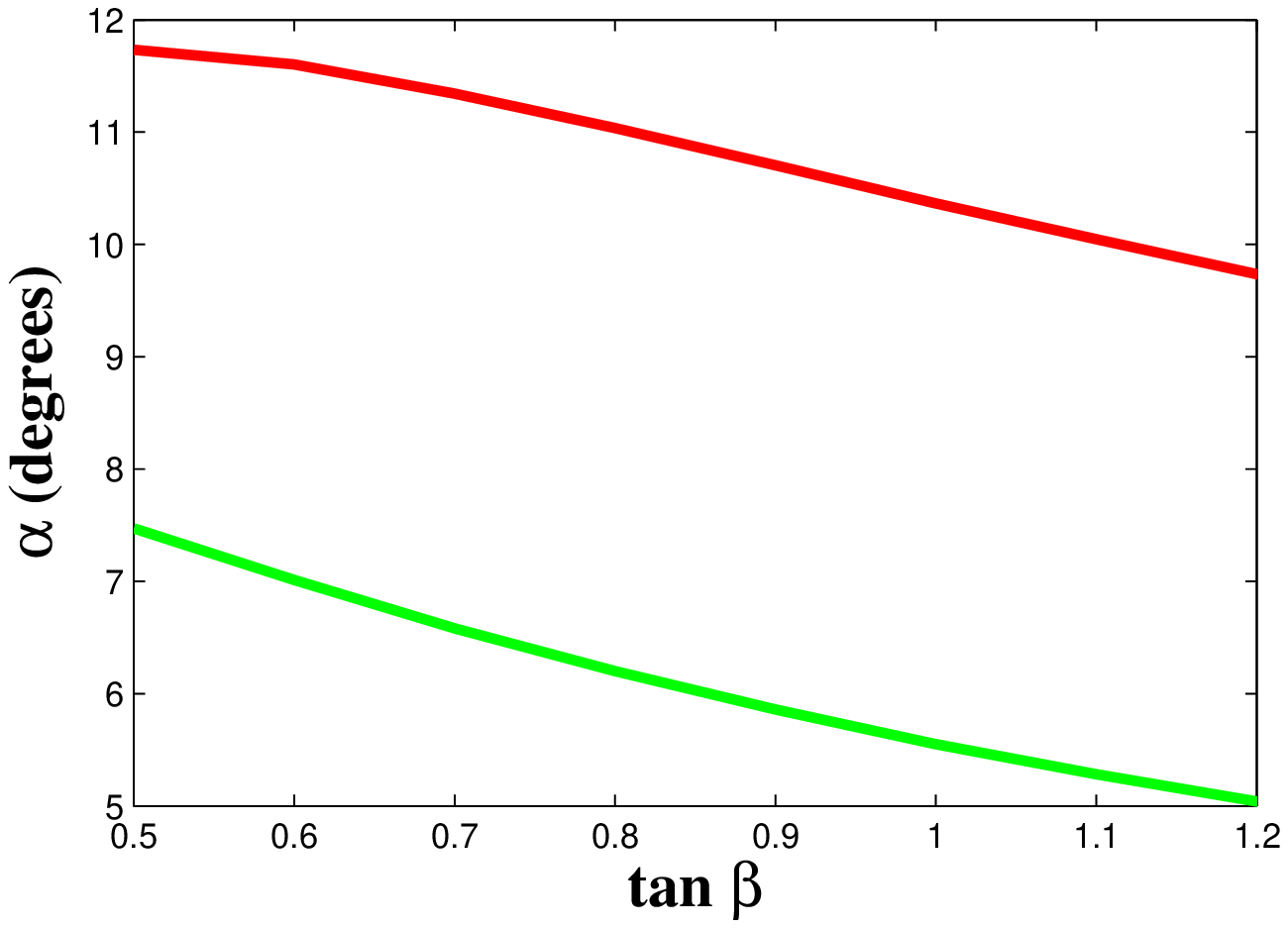}
\end{center}
\caption{\emph{The minimal value of $m_h$ (upper plot)
and the corresponding values of
the Higgs mixing angle $\alpha$ (lower plot), obtained
by choosing $\lambda_\ell(\Lambda)=0$ (see text),
as a function
of $\tan\beta$ for $m_{A} = 180$ and $250$ GeV
and $m_{q^\prime} = 600$ GeV.
See also caption to Fig.~\ref{fig2}.}
\label{fig3}}
\end{figure}

In Fig.~\ref{fig3} we plot
the minimal values of the lightest Higgs mass
$m_h$ (i.e., the solutions obtained with $\lambda_\ell(\Lambda)=0$), as a function of
$\tan\beta$, for $m_{q^\prime} = 400$ GeV and
$m_A=180,~200$ and 250 GeV.
We note that the dependence of our results
on $m_{q^\prime}$, or equivalently on $\Lambda(m_{q^\prime})$,
is negligible
for values in the range
$400 ~{\rm GeV} \lsim m_{q^\prime} \lsim 600 ~{\rm GeV}$,
as long as we choose the same boundary condition
for $\lambda_\ell(\Lambda)$, i.e.,
at the compositeness scale.

Thus, we see that $m_h \sim 125$ GeV is obtained in the h4G2HDM
with $\tan\beta \lsim 0.7$ and $m_A \lsim 250$ GeV.
This requires
small values of $\lambda_\ell(\Lambda)$ (see Fig.~\ref{fig2}) and a small
Higgs mixing angle between the
fundamental and the composite Higgs states
of a typical size $\alpha \sim {\cal O}(10^0 )$ (see Eq.~\ref{mixing} and Fig.~\ref{fig3}).
In particular, the light 125 GeV Higgs state in our model is mostly the ``fundamental"
field, while the heavy CP-even Higgs is mostly a composite state.
For the heavier Higgs we find that $m_H \sim 500 \pm 100$ GeV
within the phenomenologically viable range of values of
$m_A$ and $\tan\beta$, which give $m_h \sim 125$ GeV.

We also depict in Figs.~\ref{fig2} and \ref{fig3}
the ``exact" results, which are
obtained from a full RGE analysis including
the Yukawa couplings of the 4th generation leptons
$\ell^\prime = (\nu^\prime, \tau^\prime)$
(assuming that $\nu^\prime$ and
$\tau^\prime$ also couple to the auxiliary/composite
field $\Phi_h$ and using $m_{\ell^\prime}=200$ GeV at
the EW-scale),
of the top and of the bottom quarks,
as well as the gauge couplings.
These agree quite well with our approximate calculations.
Indeed, the observed slight
shift from
the approximate solutions is caused
mainly by ``turning on" the
Yukawa couplings of the 4th generation
leptons, and is within the estimated errors.
We note, however, that the results
are insensitive to the exact value of $m_{\ell^\prime}$,
so long as it is within the range
${\cal O}(m_W) < m_{\ell^\prime} < {\cal O}(m_{q^\prime})$.

\section{Phenomenology: Higgs signals and constraints}

Before concluding, let us briefly discuss the
collider phenomenology and the constraints
from EW precision data (EWPD) on our model. As mentioned earlier, at the EW-scale,
our model has the same spectrum and dynamics as the 4G2HDM of \cite{4G2HDM1}.
We, therefore, apply
the constraints from EWPD using the analysis in \cite{4G2HDM1,4G2HDM2},
and we update the fit made in \cite{4G2HDM4} to the 125 GeV Higgs
signals in this model.
In particular, let us define the signal strength, which is the quantity usually being used for
comparison between the measured and calculated (in a given model) Higgs signals:
 \begin{eqnarray}
 \mu^{h4G2HDM(Exp)}_{XX}=\frac{\sigma \left(pp \to h \to XX\right)_{h4G2HDM(Exp)}}{\sigma \left(pp \to h \to XX\right)_{SM}} ~,
 \end{eqnarray}
 and the individual pulls:
 \begin{eqnarray}
 {\rm Individual~pull~(XX):} ~~ \frac{\mu^{h4G2HDM}_{XX} - \mu^{Exp}_{XX}}{\sigma_{XX}} \label{IP} ~,
 \end{eqnarray}
where for the observed ratios of cross-sections, i.e., the signal strengths $\mu^{Exp}_{XX}$,
and the corresponding errors $\sigma_{XX}$, we use the latest results given in \cite{Higgssig}.

\begin{figure}[htb]
\begin{center}
\epsfig{file=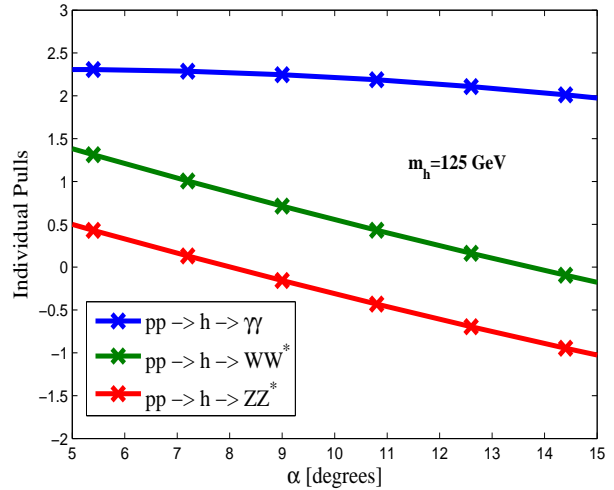,height=7cm,width=9cm,angle=0}
\vspace{-0.8cm}
\end{center}
\caption{\emph{The individual pulls as defined in Eq.~\ref{IP},
in the channels $pp \to h \to WW^*,~ZZ^*,~\gamma\gamma$, as a function of mixing angle $\alpha$ and for:
$tan \beta=1$, $m_{t^\prime}=m_{b^\prime} = 600$ GeV, $m_{l_4}=m_{\nu_4} = 200$ GeV,
$m_{H^+} \geq 200$ GeV (see text), and optimizing
with respect to $\epsilon_t$ (see text).
All points are in agreement with the EWPD.} \label{HiggsEwpd}}
\end{figure}

In Fig.~\ref{HiggsEwpd} we plot the individual pulls in the h4G2HDM
for the three most sensitive Higgs decay channels $pp \to h \to WW^*,~ZZ^*,~\gamma\gamma$,
as a function of the Higgs mixing angle $\alpha$ in the range applicable to the
h4G2HDM setup (see Fig.~\ref{fig3}) and for $\tan\beta=1$. 
In Table \ref{tab1} we further list the resulting individual pulls 
in these three Higgs decay channels for values of $\tan\beta$ around 
$\tan\beta = 1$ and for $\alpha=10^0,~20^0$. 
The rest of our model's parameter space is chosen such that:
\begin{enumerate}
\item It is consistent with EWPD:
imposing the constraints from the oblique parameters S and T,
from $Z \to b \bar b$ and from B-physics
($b \to s \gamma$ and $B-\bar B$ mixing),
following the analysis in \cite{4G2HDM1,4G2HDM2}.
\item It is consistent with the results obtained above for
our hybrid DEWSB scenario, i.e., with $m_{q^\prime} \sim {\cal O}(500)$ GeV,
$t_\beta \sim {\cal O}(1)$, $m_{H} \sim {\cal O}(500)$ GeV and
$m_{A} = 200 - 300$ GeV (see above).
\end{enumerate}

In particular, the individual pulls in Fig.~\ref{HiggsEwpd} 
and Table~\ref{tab1}
are calculated for the representative value 
$m_{t^\prime}=m_{b^\prime} = 600$ GeV, 
where, for each value of $\alpha$ and $\tan\beta$, 
we also optimize
(i.e., minimizing the $\chi^2$ for all Higgs channels)
with respect to the
parameter $\epsilon_t$,
which, in our model, represents the mixing between the 4th generation and the
3rd generation quarks (see also \cite{4G2HDM1}).
Furthermore:
\begin{itemize}
\item The individual pules are calculated for $m_{H^+}=200$ GeV,
but we have checked that they remain almost unchanged for larger $m_{H^+}$ values.
\item We assume that the Higgs is produced 100\% through the 1-loop
gluon-fusion mechanism.
\item For the calculation of the Higgs production and decay vertices
we use the latest version of Hdecay \cite{Hdecay}, with
recent NLO contributions which also
include the heavy 4th generation fermions,
where we have inserted all the relevant couplings of our model.
\item For the heavy lepton masses involved in the loops of the decays
$h \to \gamma \gamma$ and in the 1-loop NLO corrections for the cases
$h \to ZZ^*,WW^*$,
we have used $m_{l^\prime}=m_{\nu^\prime} = 200$ GeV.
\end{itemize}

We see, that a good agreement with both the Higgs data and EWPD is obtained for
$\alpha \sim 10^0$ and $\tan\beta \sim 1$ (we obtain $\chi^2 \sim 5$), 
as required for the 125 GeV hybrid Higgs state. As can be seen from 
Table~\ref{tab1}, our hybrid light Higgs is in very good agreement with the Higgs measurements 
in the $WW$ and $ZZ$ channels also for smaller (and larger) values of $\tan\beta$.    
On the other hand, there is a tension with the measurement in $pp \to h \to \gamma \gamma$ channel 
for $\tan\beta < 1$. It should be noted, however, that the 1-loop $h \to \gamma \gamma$ 
vertex is expected to be sensitive to non-decoupling contributions of the new physics 
at the compositeness scale, which we cannot estimate without knowing the details of the 
physics at that scale. In this respect, the values of $\alpha$ and $\tan\beta$ (and 
for that matter also of the value of the compositeness scale $\Lambda$) 
predicted by our hybrid EWSB mechanism should not be taken at face value
but rather as a guide (see also comment in the summary).

\begin{table}[htb]
\begin{center}
\begin{tabular}{c|c|c|c|c|c|c|c}
\multicolumn{8}{c}{individual pulls for $pp \to h \to WW^\star$} \\
\hline
$\tan\beta \Rightarrow$  & 0.7 & 0.8 & 0.9  & 1 & 1.1 &
  1.2 & 1.3 \\
\hline \hline
$\alpha=10^0$ & -0.3 & 0.04 & 0.3  & 0.5 & 0.7 & 0.8 & 0.8 \\
$\alpha=20^0$ & -1.8 & -1.6 & -1.5  & -1.5 & -1.4 & -1.4 & -1.5 \\
\end{tabular}
\begin{tabular}{c|c|c|c|c|c|c|c}
\multicolumn{8}{c}{individual pulls for $pp \to h \to ZZ^\star$} \\
\hline
$\tan\beta \Rightarrow$  & 0.7 & 0.8 & 0.9  & 1 & 1.1 & 1.2 & 1.3 \\
\hline \hline
$\alpha=10^0$ & -1.1 & -0.7 & -0.4  & -0.2 & -0.01 & 0.1 & 0.2 \\
$\alpha=20^0$ & -2.8 & -2.6 & -2.5  & -2.4 & -2.4 & -2.4 & -2.5 \\
\end{tabular}
\begin{tabular}{c|c|c|c|c|c|c|c}
\multicolumn{8}{c}{individual pulls for $pp \to h \to \gamma \gamma$} \\
\hline
$\tan\beta \Rightarrow$  & 0.7 & 0.8 & 0.9  & 1 & 1.1 & 1.2 & 1.3 \\
\hline \hline
$\alpha=10^0$ & 73.4 & 34.2 & 13.3  & 1.9 & -2.7 & -2.4 & -1.9 \\
$\alpha=20^0$ & 7.1 & -2.9 & -7.0  & -6.7 & -6.6 & -6.8 & -7.1 \\
\end{tabular}
\caption{\emph{The individual pulls as defined in Eq.~\ref{IP},
in the channels $pp \to h \to WW^*,~ZZ^*,~\gamma\gamma$, 
for $\alpha=10^0$ and $20^0$ and for values of 
$\tan\beta$ in the range $\tan\beta = 1 \pm 0.3$. 
The rest of the parameters are as in Fig.~\ref{HiggsEwpd} 
and all points are in agreement with the EWPD.}}
\label{tab1}
\end{center}
\end{table}

Finally, we note that our model can also
lead to interesting new signatures of
the heavy quarks, such as the
flavor changing decay $t^\prime \to t h$,
which can be searched for in the present LHC data
using a specific search strategy, see
\cite{4G2HDM3}.
Indeed, very recently the ATLAS collaboration has performed
a search for $t^\prime$ of an SU(2) doublet,
using selection criteria designed for the case in which
$t^\prime \to t h$ followed by
$h \to \bar b b$ \cite{newATLAS}.
They obtained a new bound
$m_{t^\prime} \gsim 800$ GeV, which is
appreciably stronger than the previously existing
bound (see \cite{olderATLAS}).
This new bound is in tension
with our DEWSB framework,
since such a heavy 4th generation doublet, if it is chiral,
will be non-perturbative at energy scales
already smaller than its mass.$^{[6]}$\footnotetext[6]{
We note, however, that the recent search performed by ATLAS
in \cite{newATLAS} assumes $t^\prime \to t h$ followed by
$h \to b \bar b$, while in our low energy h4G2HDM,
$h \to gg$ might be the dominant decay, see \cite{4G2HDM4}.}
We emphasized, however, that
our hybrid DEWSB framework can be equally constructed
with TeV-scale vector-like quarks \cite{vectorquarks}.
In such a case, our hybrid
mechanism can be applied
also with a higher compositeness scale
of several TeV, without
being in conflict with perturbativity of the
vector-like quarks Yukawa couplings.
We leave these issues for a future work.

\section{Summary}

To summarize, we have constructed a hybrid scenario for
DEWSB, where both a fundamental-like
Higgs field
and a condensate of a strongly coupled heavy quark
sector participate in EWSB. This yields a partly dynamical EWSB setup,
and the resulting low-energy theory is a hybrid 2HDM with
another family of heavy fermions.
In particular, in this model
one Higgs (mostly fundamental) couples only to the light SM fermions while the 2nd
Higgs (mostly composite) couples only to the heavy (dynamical) new fermions.
The proposed DEWSB framework results
in a compositeness scale $\Lambda \sim $ TeV
and a phenomenologically viable low energy  setup,
which closely resembles the recently proposed model of \cite{4G2HDM1}
and, as we show, is consistent with current EW
precision and flavor data.
In particular,
in spite of the heaviness of the dynamical quarks,
$m_{q^\prime} \sim {\cal O}(500)$ GeV,
and the resulting low TeV-scale threshold for the strong dynamics,
a viable Higgs candidate is obtained, which has a mass
$m_h \sim 125$ GeV and
properties {\it very similar} to the SM Higgs,
when $\tan\beta \sim {\cal O}(1)$,
$m_A \sim m_{H^+} \sim 200 - 300$ GeV and $m_H \sim {\cal O}(500)$ GeV.
Indeed, we perform a fit to the recently measured 125 GeV Higgs signals,
also imposing the constraints from precision EW data, and demonstrate
the consistency of our model with all current collider phenomenology.

We emphasize that the purpose of this work is
to present a mechanism that addresses the low-energy outcome
of a possible TeV-scale strong sector and not to discuss the details of the
strongly interacting theory. Indeed, we show that this mechanism
(the hybrid structure of DEWSB) is very predictive without knowing
the details of the physics at the cutoff scale.
In this respect, the exact
values obtained for the free-parameters of the model,
i.e., for $m_A$, $\tan\beta$ and $\Lambda$,
have no deep meaning, but should rather be interpreted
as a guide.
For example, the exact position of a Landau pole cannot be taken at
face value, since the threshold effects are unknown
without knowing the details of the UV completion
of the model. In particular,
the expression we get for $\Lambda$ in Eq.~\ref{lambda} is an estimate
from the RGE of the model and it is valid only in the
range where the RGE are valid and can be used in a perturbative manner.
Clearly, this is not the case as one approaches the compositeness scale.
For that reason it is fair to say that $\Lambda \sim 1$ TeV
should be rather viewed as $\Lambda \sim$ few TeV. Similarly,
$\tan\beta \sim 0.7$ should be interpreted as
$\tan\beta \sim {\cal O}(1)$ and the value we get for the
Higgs mixing angle, i.e., $\alpha \sim {\cal O}(10^0)$
should be taken as a guide.

\bigskip

{\bf Acknowledgments:} SBS and MG acknowledge research support from the Technion.
The work of AS was supported in part by DOE contract
\#DE-AC02-98CH10886(BNL).


\newpage

\begin{thebibliography}{99}

\bibitem{LHCHiggs} S. Chatrchyan {\it et al.}, [CMS Collaboration],
Phys. Lett. {\bf B716}, 30 (2012);
G. Aad {\it et al.}, [ATLAS Collaboration], Phys. Lett. {\bf B716}, 1 (2012).

\bibitem{EWSB-papers} S. Weinberg, Phys. Rev. {\bf D19}, 1277 (1979);
L. Susskind, Phys. Rev. {\bf D20}, 2619 (1979). See also,
R. Jackiw and K. Johnson, Phys. Rev. {\bf D8}, 2386 (1973);
J.M. Cornwall and R.E. Norton, Phys. Rev. {\bf D8}, 3338 (1973).

\bibitem{Nambu} Y. Nambu and G. Jona-Lasinio Phys. Rev. {\bf 122},
345 (1961).

\bibitem{Bardeen} V.A. Miransky, M. Tanabashi and K. Yamawaki, Phys. Lett. {\bf B221}, 177 (1989);
{\it ibid.}, Mod. Phys. Lett. {\bf A4}, 1043 (1989);
W.A. Bardeen, C.T. Hill and M. Lindner, Phys. Rev. {\bf D41}, 1647 (1990);
see also Y. Nambu in e.g., EFI report \#89-08, 1989 (unpublished).

\bibitem{DEWSB-Pseudo} G. Burdman, C.E.F. Haluch, JHEP {\bf 1112}, 038 (2011);
B. Holdom, Phys. Lett. {\bf B721}, 290 (2013).

\bibitem{Hill1994} C.T. Hill, Phys. Lett. {\bf B345}, 483 (1995).

\bibitem{newpaper} A. Smetana, Eur. Phys. J. {\bf C73}, 2513 (2013).

\bibitem{DEWSB-heavyF} B. Holdom, Phys. Rev. Lett. {\bf 57}, 2496 (1986),
Erratum-ibid. {\bf 58}, 177 (1987);
S.F. King, Phys. Lett. {\bf B234}, 108 (1990);
P.Q. Hung and G. Isidori Phys. Lett. {\bf B402}, 122 (1997);
B. Holdom, JHEP {\bf 0608}, 76 (2006);
Y. Mimura, W.S. Hou and H Kohyama,  JHEP {\bf 1311}, 048 (2013).

\bibitem{luty} M.A. Luty, Phys. Rev. {\bf D41}, 2893 (1990).

\bibitem{DEWSB-HLP} C. Hill, M. Luty and E.A. Paschos, Phys. Rev. {\bf D43}, 3011 (1991).

\bibitem{burdman1} G. Burdman and L. Da Rold, JHEP {\bf 0712}, 86 (2007).

\bibitem{hunghybrid} P.Q. Hung, C. Xiong, Nucl. Phys. {\bf B848}, 288 (2011);
for an earlier discussion see also,
``Dynamical symmetry breaking due to strong coupling Yukawa interaction"
by M. Tanabashi, K. Yamawaki and K. Kondo,
in {\it Nagoya 1989, Proceedings, Dynamical symmetry breaking} 28-36.



\bibitem{DEWSB-multiH}
G. Burdman, L. Da Rold, O. Eboli and R.D. Matheus,
Phys. Rev. {\bf D79}, 075026 (2009);
M. Hashimoto and V.A. Miransky, Phys. Rev. {\bf D81}, 055014 (2010);
A.E.C. Hernandez, C.O. Dib, H.N. Neill and A.R. Zerwekh, JHEP {\bf 1202}, 132 (2012);
P.Q. Hung, C. Xiong, Nucl. Phys. {\bf B847}, 160 (2011);
P.Q. Hung, C. Xiong, Phys. Lett. {\bf B694}, 430 (2011);
C.M. Ho, P.Q. Hung and T.W. Kephart, JHEP {\bf 1206}, 45 (2012).
G. Burdman, L. de Lima and R.D. Matheus, Phys. Rev. {\bf D83}, 035012 (2011).

\bibitem{hybrid-top} B. Chung, K.Y. Lee, D.-W. Jung and P. Ko, JHEP {\bf 0605}, 010 (2006).

\bibitem{vectorquarks} See e.g.,
N. Arkani-Hamed, A.G. Cohen, H. Georgi, Phys. Lett. {\bf B513}, 232 (2001);
N. Arkani-Hamed, A. Cohen, E. Katz, A. Nelson, JHEP {\bf 0207}, 034 (2002);
M. Perelstein, Prog. Part. Nucl. Phys. {\bf 58}, 247, 2007);
R. Contino, L.D. Rold, A. Pomarol, Phys. Rev. {\bf D75}, 055014 (2007);
M. Carena, E. Ponton, J. Santiago, C.E. Wagner, Nucl. Phys. {\bf B759}, 202 (2006); J.A. Aguilar-Saavedra, JHEP {\bf 11}, 030 (2009).

\bibitem{4G2HDM1} S. Bar-Shalom, S. Nandi and A. Soni, Phys. Rev. {\bf D84}, 053009 (2011).

\bibitem{SM4-excluded} For a recent analysis see, O. Eberhardt {\it et. al.},
Phys. Rev. Lett. {\bf 109}, 241802 (2012).


\bibitem{4G2HDM2} S. Bar-Shalom, M. Geller, S. Nandi and A. Soni,
arXiv:1208.3195 [hep-ph], a review published in
a special issue of Advances in High Energy
Physics (AHEP) on ``{\it Very Heavy Quarks at the LHC}",
{\bf Adv.High Energy Phys.2013}, 672972 (2013).

\bibitem{4G2HDM3} M. Geller, S. Bar-Shalom and G. Eilam, Phys. Lett. {\bf B715}, 121 (2012).

\bibitem{4G2HDM4} M. Geller, S. Bar-Shalom, G. Eilam and A. Soni,
Phys. Rev. {\bf D86}, 115008 (2012).

\bibitem{4G2HDMpheno}
M. Hashimoto, Phys. Rev. {\bf D81}, 075023 (2010);
H.-S. Lee and A. Soni, Phys. Rev. Lett. {\bf 110}, 021802 (2013);
M. Sher, Phys. Rev. {\bf D61}, 057303 (2000);
L. Bellantoni, J Erler, J. Heckman and E. Ramirez-Homs,
Phys. Rev. {\bf D86}, 034022 (2012);
W. Bernreuther, P. Gonzales and M. Wiebusch, Eur. Phys. J. {\bf C69}, 31 (2010);
John F. Gunion, arXiv:1105.3965 [hep-ph];
X.-G. He and G. Valencia, Phys. Lett. {\bf B707}, 381 (2012);
N. Chen and H. He, JHEP {\bf 1204}, 062 2012;
E. De Pree, G. Marshall and M. Sher, Phys. Rev. {\bf D80}, 037301 (2009);
M. S. Chanowitz, arXiv:1212.3209 [hep-ph].

\bibitem{Hill1991} C.T. Hill, Phys. Lett. {\bf B266}, 419 (1991).

\bibitem{PQ1} ``Top mode standard model" by K. Yamawaki,
in {\it Strong Coupling Gauge Theories and beyond}
(SCGT 90), Proceedings 28-31, Jul 1990, Nagoya, Japan.

\bibitem{PQ} R.D. Peccei and H.R. Quinn, Phys. Rev. {\bf D16}, 1791 (1977).

\bibitem{branco} G.C Branco et. al., Phys. Rept. {\bf 516} (2012).

\bibitem{Higgssig} See e.g., ``Higgs bosons in the SM and beyond",
talk given by G. Landsberg at {\it The 2013 European Physical Society Conference on High Energy Physics}, EPSHEP 2013, Stockholm, Sweden, 18-24 July, 2013.

\bibitem{Hdecay} A. Djouadi, J. Kalinowski and M. Spira,
Comput. Phys. Commun. {\bf 108}, 56 (1998), with contributions
from arXiv:hep-ph/9704448 [hep-ph].

\bibitem{newATLAS} ATLAS Collaboration,
ATLAS-CONF-2013-018, ATLAS-COM-CONF-2013-024.

\bibitem{olderATLAS} Georges Aad (Freiburg U.) {\it et al.} (ATLAS Collaboration),
Phys. Lett. {\bf B718}, 1284 (2013).

\end{thebibliography}
\end{document}